# Breaking Limits of Line-of-Sight MIMO Capacity in 6G Wireless Communications

Haiyue Jing, *Student Member, IEEE*, Wenchi Cheng, *Senior Member, IEEE*, and Wei Zhang, *Fellow, IEEE*

## Abstract

Multiple-input-multiple-output (MIMO) has been proved its success for the fourth generation (4G) long term evolution (LTE) and is one of the key technical enablers for evolved mobile broadband (eMBB) in the fifth generation (5G) wireless communications. However, along with the number of antennas eventually increased to be extremely large and one-hop communication distance gradually reduced, how to significantly increase the capacity for line-of-sight (LOS) MIMO becomes more and more urgent. In this article, we introduce the quasi-fractal uniform circular array (QF-UCA) antenna structure based MIMO wireless communications, which can adequately exploit the potential of MIMO in LOS channel and greatly increase the capacity with low complexity demodulation schemes. Specifically, three advantages regarding QF-UCA based LOS MIMO are reviewed. Then, research challenges on transceiver alignment, low-rank channel matrix, extended dimensions of QF-UCA, maximum number of orthogonal streams, and the corresponding potential solutions are discussed. Compared with traditional scattering-depended MIMO communications, the QF-UCA based LOS MIMO wireless communication can achieve high-efficient transmission in LOS channel.

Haiyue Jing and Wenchi Cheng are with the State Key Laboratory of Integrated Services Networks, Xidian University, Xi'an, 710071, China (e-mails: hyjing@stu.xidian.edu.cn and wccheng@xidian.edu.cn).

Wei Zhang is with the School of Electrical Engineering and Telecommunications, University of New South Wales Sydney, NSW, Australia (e-mail: wzhang@ee.unsw.edu.au).

## I. INTRODUCTION

Due to its main advantage of capacity enhancement, multiple-input-multiple-output (MIMO) has been widely investigated and used in the fourth generation (4G) and the fifth generation (5G) wireless communications [1] as well as become an even-increasingly important part in wireless standardization. As wireless communications migrate from 5G to the sixth generation (6G), MIMO is expected to further increase the capacity of wireless communications for meeting the requirements of the high speed transmission based evolved mobile broadband (eMBB) evolution [2], [3].

To increase the capacity of wireless communications with MIMO, more and more antennas are equipped at the transmitter and receiver to increase the orthogonal streams and relative high frequency resources are explored to broaden the transmission frequency bands [4], [5]. However, the increasing number of antennas generally impacts the correlation of channels while the increasing high frequency bands limits the transmission distance, which impose urgent but important requirements for MIMO transmission in line-of-sight (LOS) channel [6], [7]. LOS MIMO transmission is a typical scenario in point-to-point communications and millimeter-wave (mmWave) communication.

By properly placing transmit antennas and receive antennas, uniform circular array (UCA) is with the potential to increase the orthogonal streams and capacity for LOS MIMO wireless communications [8], [9]. Moreover, UCA can achieve low-complexity but high performance demodulation with the linear detection scheme as well as the maximum-likelihood (ML) detection scheme [10], relying on that the channel matrix of LOS MIMO is circulant. However, existing researches on maximum orthogonal streams are restricted to the number of array-elements for UCA based LOS MIMO wireless communications, i.e., the maximum number of orthogonal streams is restricted to the array-element number of UCA antenna [11], [12]. It is now very urgent and highly expected to break through the limit of the array-element number to increase orthogonal streams, thus achieving high capacity transmission. Until now few studies reported on how to adequately achieve multiple streams exceeding the number of array-elements corresponding to the UCA, which is beyond the traditional concept of scattering-based multiple antennas wireless communications.

To break through the limit of the array-element number, in this article we introduce the quasi-fractal UCA (QF-UCA) antenna, which is a new geometry layout of antenna and can be used to achieve more orthogonal streams, i.e., the number of orthogonal streams is larger than that of array-elements. We also give different examples about QF-UCA antennas. Then, we propose the QF-UCA based LOS MIMO wireless communication and use the geometric axisymmetry of circular arrays together with the fractal geometry principle to implement spatial multiplexing in a new way for LOS MIMO wireless communications. Next, we introduce three advantages of the QF-UCA based LOS MIMO wireless communications, including high capacity transmission, low-complexity demodulation, as well as easy and accurate channel state information (CSI) estimation. Numerical analyses are given to further demonstrate advantages. The challenges regarding QF-UCA based LOS MIMO wireless communications are then discussed, including transceiver alignment, low-rank channel matrix, extended dimensions of QF-UCA, and maximum number of orthogonal streams. Some potential solutions corresponding to the challenges are also discussed.

The rest of article is organized as follows. Section II introduces the QF-UCA based LOS MIMO wireless communication with what is it and how the QF-UCA based LOS MIMO works. Section III shows the advantages of QF-UCA based LOS MIMO wireless communications. Section IV discusses its main challenges and potential solutions. Finally, the conclusion is given in Section V.

## II. The QF-UCA Based LOS MIMO Wireless Communications

Figure 1 depicts the QF-UCA based LOS MIMO wireless communication system, where both transmitter and receiver are with QF-UCA antennas. QF-UCA is a two-dimensional array antenna architecture consisting of the basic inner-UCA composed of array-elements and the inter-UCA composed of inner-UCAs. For the QF-UCA based MIMO wireless communication system as shown in Fig. 1, the transmitter contains inner-UCA modulation, inter-UCA modulation, power combiner, and transmit QF-UCA antenna. The inner-UCA modulation and the inter-UCA modulation form the two-dimension inverse discrete Fourier transform (IDFT) for input signals. The power combiner along with transmit QF-UCA antenna are responsible for integrating and emitting multiple signals. The receiver contains receive QF-UCA antenna, power splitter, inter-UCA demodulation, post-decoding, inner-UCA demodulation, and multiple branch signal detection. The receive QF-UCA antenna and power splitter are responsible for receiving and separating multiple signals. The inter-UCA demodulation, the inner-UCA demodulation, and the post-decoding are used to keep the signals orthogonal.

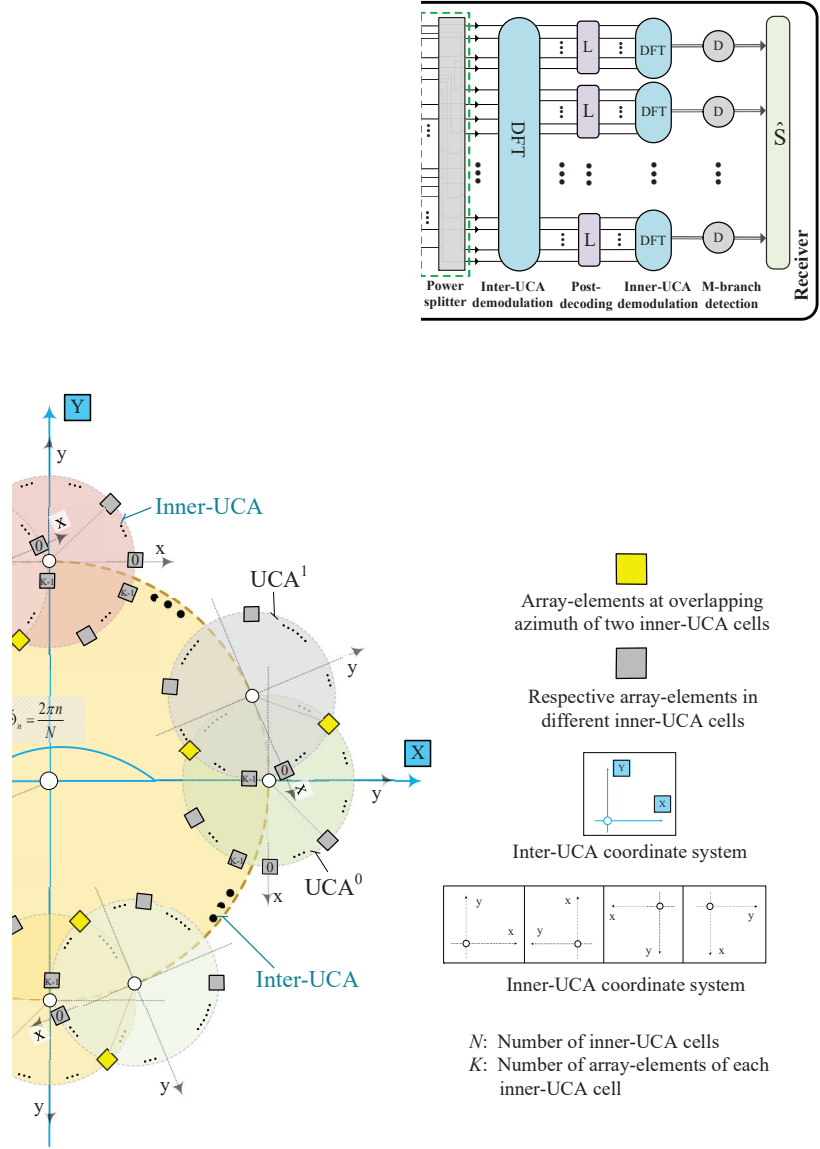

Fig. 1. The system model for the QF-UCA based LOS MIMO wireless communication.

As illustrated in Fig. 1, the QF-UCA antenna contains $N$ inner-UCAs and each inner-UCA equipping $K$ array-elements. For the inner-UCA coordinate system, x-axis is set as the direction from the center of inner-UCA cell to the first array-element while z-axis is the centering normal line pointing to the directly opposite receive inner-UCA cell. For the inter-UCA coordinate system, x-axis is set as the direction from the center of inter-UCA to the first inner-UCA cell named UCA$^0$ while z-axis is the centering normal line pointing to the receive QF-UCA antenna. Correspondingly, y-axis are decided by the right-hand spiral rule. Based on this, the position of each array-element in three-dimension (3D) space, which is related to the azimuth of the array-elements within the inner-UCA cell and the azimuth of the inner-UCA cells, can be described by the two coordinates as shown in Fig. 1.

## A. Design for QF-UCA Antenna

The key of designing QF-UCA antenna is the array-elements sharing for the inner-UCA cells. The shared array-elements can facilitate to achieve more orthogonal streams exceeding the number of array-elements. Regarding different kinds of placement for array-elements, different cases about the layout of QF-UCA are given in the following.

**Case 1: No shared array-element between two adjacent inner-UCA cells.** For the case with no shared array-element between any two adjacent inner-UCA cells as shown in Fig. 2, the number of orthogonal streams is equal to the number of the array-elements and they are $NK$.

**Case 2: One shared array-element between two adjacent inner-UCA cells.** For the case with one shared array-element between two adjacent inner-UCA cells and the radius of the inner-UCA is smaller than that of the inter-UCA as shown in Fig. 2, the number of orthogonal streams is $NK$ and the number of array-elements in the QF-UCA is $N(K-1)$.

**Case 3: Two shared array-element between two adjacent inner-UCA cells.** In this case, there are two shared array-elements between two adjacent inner-UCA cells and there is no shared array-element between two non-adjacent inner-UCA cells as shown in Fig. 2. The number of orthogonal streams is $NK$ and the number of array-elements in the QF-UCA is $N(K-2)$.

**Case 4: Two shared array-elements between two non-adjacent inner-UCA cells.** When there exist shared array-elements arranged for $M$ adjacent inner-UCA cells as shown in Fig. 2, the number of orthogonal streams is $NK$ and the minimum number of array-elements in the QF-UCA is $N(K-M)$.

**Case 5: One shared array-element among all inner-UCA cells.** When the radius of the inner-UCA cell is equal to that of the inter-UCA and there is one shared array-element arranged for all inner-UCA cells as shown in Fig. 2, the number of orthogonal streams is $NK$ and the number of array-elements in the QF-UCA is $N(K-2)+1$.

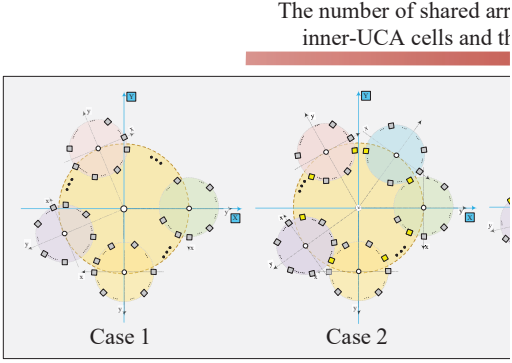

Fig. 2.   Different cases about the layout of QF-UCA antenna.

Observing Fig. 2, it is clear that the total number of array-elements for the QF-UCA decreases as the number of shared array-elements increases with fixed number of orthogonal streams. Generally, the maximum number of orthogonal streams is $2N$ more than the number of array-elements in the QF-UCA where there are shared array-elements between any two adjacent inner-UCA cells and there is no shared array-element between any two non-adjacent inner-UCA cells. Since the radius of the inter-UCA cell is not smaller than that of the inner-UCA as shown in Fig. 2, the radius of the inter-UCA for the QF-UCA in case 5 is smaller than those in other cases.

### B. The QF-UCA Based MIMO Transmission

At the transmitter, signals in the same branch are first given a specified phase gradient which is related to the phase order used for the inner-UCA (the first-dimension) modulation and the azimuth difference of any two adjacent array-elements within the inner-UCA cell. This process is equivalent to applying the $K$-point IDFT for the $K$ input signals. The inter-UCA (the second-dimension) modulation is performed that the signals corresponding to the same inner-UCA cell are with a phase gradient which is the product of the phase order used for the inter-UCA modulation and the azimuth difference of any two adjacent inner-UCA cells. This process is equal to the $N$-point IDFT for $K$ signal vectors. Thus, the modulation at the transmitter can be considered the two-dimension IDFT for the input signals. At the receiver, received signals are first split into multiple branches. Then, the first-dimension demodulation is to separate the signals

TABLE I

Complexity Comparison for LOS MIMO Systems With Different Antennas

| MIMO systems | The number of complex additions | The number of complex multiplications |
|---|---|---|
| MIMO systems with QF-UCA | $NK \log_2(NK) + NKV$ | $\frac{NK}{2} \log_2(NK) + NK(V+1)$ |
| MIMO systems with UCA | $NK \log_2(NK) + NKV$ | $\frac{NK}{2} \log_2(NK) + NK(V+1)$ |
| MIMO systems with ULA | $N^2 K^2 V^{NK}$ | $N^2 K^2 + NKV^{NK}$ |

from inter-UCA with performing phase compensation based on the azimuths of receive inner-UCA cells, which is the process of discrete Fourier transform (DFT) for the signal vectors. Then, the signals are demodulated without interference using the second-dimension demodulation. Using the modulation and demodulation schemes, more orthogonal streams can be achieved, i.e., the number of orthogonal streams is larger than that of array-elements, which leads to high capacity transmission for wireless communications.

## III. Advantages for QF-UCA Based LOS MIMO Wireless Communications

Due to the fractal geometry of QF-UCA, which can achieve more orthogonal streams, and the geometric axisymmetry of QF-UCA, which makes the channel matrix circulant, QF-UCA based LOS MIMO wireless communication can achieve high capacity transmission, low-complexity demodulation, as well as easy and accurate CSI estimation. Three basic advantages regarding QF-UCA based LOS MIMO wireless communications are reviewed as follows.

**Advantage 1: High capacity transmission** — The QF-UCA antenna structure is a two-dimension extension of UCA and can achieve parallel transmissions without interference among different streams. Typically, the number of orthogonal streams is much more than the number of array-elements in the QF-UCA, which shows the superiority of QF-UCA based LOS MIMO wireless communications in the aspect of capacity enhancement. Thus, QF-UCA can be used to significantly increase the capacity of wireless communications.

Figure 3 shows two different layouts of QF-UCA antenna and their corresponding capacities,

where $U$ represents the total number of array-elements in the QF-UCA. Fig. 3 gives the case of two shared array-elements between two adjacent inner-UCA cells when $N = 4$ and $K = 4$. The corresponding total number of array-elements is 9, where 16 orthogonal streams can be achieved. Fig. 3 also gives the case of two shared array-elements between two adjacent inner-UCA cells when $N = 4$ and $K = 8$. The corresponding total number of array-elements $U$ is 25, where 32 orthogonal streams can be achieved. Then, the capacities of LOS MIMO wireless communications based on the QF-UCA antennas and the single-loop UCA antennas are depicted. We set the communication system operating at the 3G Hz frequency band and the bandwidth is set as 1M Hz. Without additional power and frequency consumption, compared with the single-loop UCAs with 9 array-elements (the legend with 9-element UCA) and 25 array-elements (the legend with 25-element UCA), the 9 array-elements QF-UCA antenna based LOS MIMO wireless communications (the legend with 9-element QF-UCA) and the 25 array-elements QF-UCA antenna based LOS MIMO wireless communications (the legend with 25-element QF-UCA) are with a huge enhancement in capacity, respectively. This is because the QF-UCA antenna based LOS MIMO wireless communications can achieve more orthogonal streams with its number exceeding the array-element number. The results show a new way to investigate more multiplexings beyond that of traditional multiple antennas wireless communication systems in correlated channel scenarios. Furthermore, the obtained capacities of 9 array-elements QF-UCA antenna and 25 array-elements QF-UCA antenna based LOS MIMO wireless communications are even larger than those of single-loop UCAs with 16 array-elements (the legend with 16-element UCA) and 32 array-elements (the legend with 32-element UCA), respectively. This is because the placements of array-elements changes the correlations for LOS MIMO wireless communications and the number of orthogonal streams increases, thus resulting in relatively high capacity.

**Advantage 2: Low-complexity demodulation** —When the transmitter and receiver equipped with QF-UCAs are aligned with each other, the signals are modulated by the two-dimension IDFT at the transmitter and demodulated by the two-dimension DFT at the receiver. Then, the signals can be considered free-interference and the complexity of detection schemes, such as linear detection scheme and ML detection scheme, can be significantly reduced while the bit error rate (BER) keep unchanged since the channel matrix is a two-dimension circulant matrix.

The computational complexities corresponding to different MIMO systems using ML detection scheme to demodulate the signals are listed in Table I, where $V$ represents the size of modulation alphabet. The numbers of complex additions and complex multiplications for LOS MIMO systems with QF-UCA antennas are the same to those for LOS MIMO with single-loop UCA antennas. The numbers of complex additions and complex multiplications for LOS MIMO systems with QF-UCA antennas are much smaller than those for LOS MIMO with uniform linear array (ULA) antennas. For example, When $N = 4$, $K = 8$, and $V = 8$, the numbers of complex additions and complex multiplications for LOS MIMO systems with ULA antennas are approximately $1.95 \times 10^{29}$ and $1.89 \times 10^{27}$ times, respectively, more than those for QF-UCA based LOS MIMO systems.

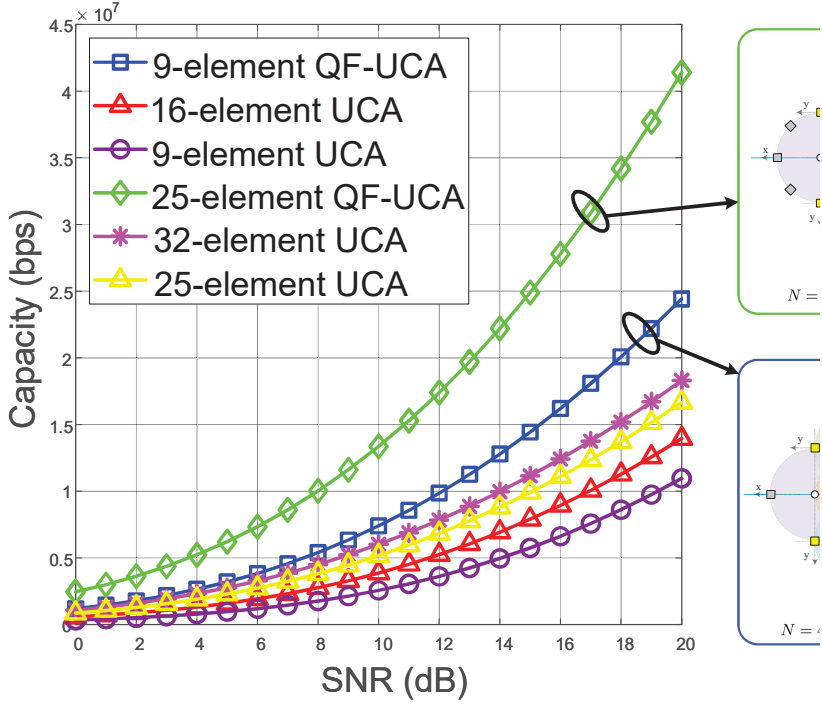

Fig. 3. Two different layouts of QF-UCA antenna and capacities of LOS MIMO with different antennas.

**Advantage 3: Easy and accurate CSI estimation** —According to the design of QF-UCA, the difference of phases can be utilized to accurately estimate the CSI because the channel matrix can be considered as a two-dimension circulant matrix. Also, using the properties of circulant matrix, the CSI estimation can easily be adaptive to practical wireless communications both in accuracy and efficiency since the CSI can be obtained by calculation.

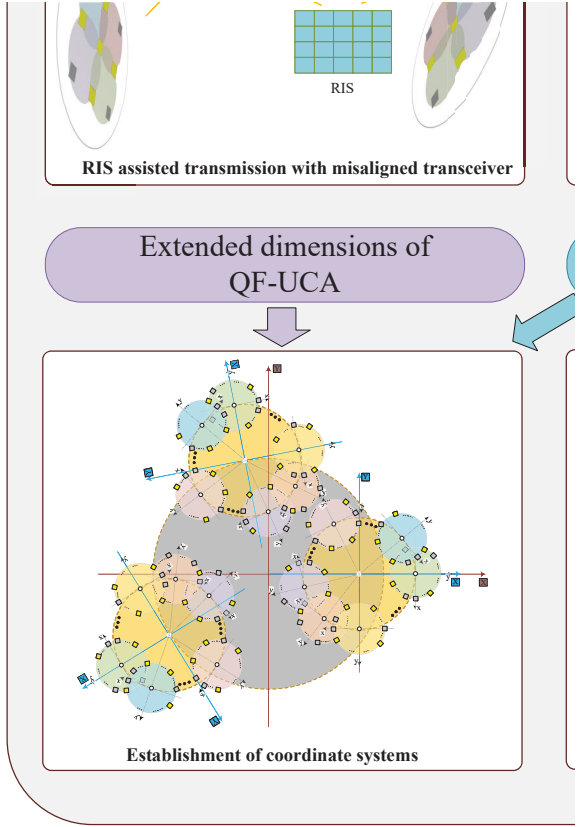

Fig. 4.  Potential solutions for the challenges of QF-UCA based LOS MIMO wireless communications.

## IV. Challenges and Future Research for QF-UCA based LOS MIMO Wireless Communications

Despite all above advantages, there are mainly four challenges regarding QF-UCA based MIMO wireless communications, including the transceiver alignment, the low-rank channel matrix, the extended dimensions of QF-UCA, and the maximum number of orthogonal streams. In the following, we discuss these challenges and give the potential solutions.

**Challenge 1: Transceiver alignment.** —In practical wireless communications scenarios, it is very difficult to ensure the strict alignment between the transmitter and the receiver. When the transmitter and receiver are misaligned, achieving high capacity with existing schemes is very difficult since there exists interference among different streams and the complexity of signal demodulation significantly increases. Some schemes have been proposed to eliminate the impact of the transceiver misalignment on the capacity. For example, a beamforming scheme with fast symbol-wise ML detection for UCA based LOS MIMO wireless communications is developed [10] and the scheme can eliminate the interference among different streams. In addition, reconfigurable intelligent surface (RIS) can be applied to the QF-UCA based MIMO wireless communications. As shown in Fig. 4, placing RISs between the transmitter and receiver and designing the phase compensation on the RISs, the impact of transceiver misalignment can be eliminated, thus achieving high capacity transmission for wireless communications [13].

**Challenge 2: Low-rank channel matrix.** —Due to the low-rank channel matrix, it is difficult to convey the data streams using all orthogonal streams for LOS MIMO wireless communications, thus decreasing the capacity for wireless communications. There are some schemes which can be used to increase the rank of channel matrix. By properly placing the transmit antennas and receive antennas according to the mutual information and designing the distance among neighboring array-elements, a high-rank channel matrix can be constructed to achieve high capacity transmission for LOS MIMO wireless communications. As shown in Fig. 4, since the scattering can facilitate to achieve high-rank channel matrix for wireless communications, RIS can also be used to increase the rank of channel matrix by designing the placement positions and the phase compensation matrices on the RISs [14], [15].

**Challenge 3: Extended dimensions of QF-UCA.** —In this article, we mainly focus on the QF-UCA based two-dimension extension of UCA. When the QF-UCA antenna structure is a multiple-dimension extension of UCA, the number of orthogonal streams is much more than the number of array-elements, thus significantly increasing the capacity for LOS MIMO wireless communications. As shown in Fig. 4, if there exist shared inner-UCAs, the establishment of coordinate systems, the derivations of the multiple-dimension channel matrix, and the multiple-dimension decomposition of signals will be very complicated. It is a good idea to make approximations with small error on the distances among different array-elements corresponding to different inner-UCAs, which can easily decompose the signal vectors regarding to the multiple-dimension QF-UCAs. Using the similarity transformation of channel matrix can also recover the signals at the cost of complexity increase.

**Challenge 4: Maximum number of orthogonal streams.** —Since the number of orthogonal streams increases as the number of shared array-elements increases, it is intriguing to increase the number of shared array-elements by increasing the number of intersected non-adjacent inner-UCAs, as shown in Fig. 4, and increasing the number of extended dimensions of QF-UCA as shown in Fig. 4. When there are lots of array-elements equipped at the transmitter and receiver, such as base station, how to design the number of intersected non-adjacent inner-UCAs and the number of extended dimensions of QF-UCA to achieve the maximum number of orthogonal streams for LOS MIMO wireless communications is a very important problem. To reveal the relationships between the number of intersected non-adjacent inner-UCAs and the number of extended dimensions of QF-UCA is an expecting idea to fight this challenge.

## V. Conclusions

In this article, we introduced the QF-UCA for achieving more orthogonal streams and we proposed the QF-UCA based LOS MIMO wireless communications to achieve high capacity wireless communications. The advantages of QF-UCA based LOS MIMO wireless communications are reviewed, including high capacity transmission, low-complexity demodulation, as well as easy and accurate CSI estimation. Four challenges and the potential solutions are also discussed. In conclusion, the number of orthogonal streams of QF-UCA based LOS MIMO wireless communications have the potential to exceed the number of array-elements in QF-UCA, creating many opportunities for future wireless communications research.

## VI. BIOGRAPHIES

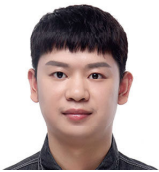

**Haiyue Jing** received the B.S. degree in telecommunication engineering from Xidian University, China, in 2017. He is currently pursuing the Ph.D. degree in telecommunication engineering at Xidian University. His research interests focus on B5G/6G wireless networks, OAM based wireless communications, and LOS MIMO wireless communications.

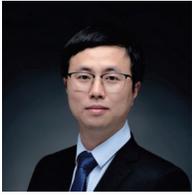

**Wenchi Cheng** received his B.S. and Ph.D. degrees in telecommunication engineering from Xidian University in 2008 and 2013, respectively, where he is a full professor. He was a visiting scholar with the Department of Electrical and Computer Engineering, Texas A&M University, College Station, from 2010 to 2011. His current research interests include B5G/6G wireless networks, emergency wireless communications, and OAM based wireless communications. He has published more than 100 international journal and conference papers in the IEEE JSAC, IEEE magazines, and IEEE transactions, and at conferences including IEEE INFOCOM, GLOBECOM, ICC, and more. He has served or is serving as an Associate Editor for the IEEE Systems Journal, IEEE Communications Letters, and IEEE Wireless Communications Letters, as the Wireless Communications Symposium Co-Chair for IEEE ICC 2022 and IEEE GLOBECOM 2020, the Publicity Chair for IEEE ICC 2019, and the Next Generation Networks Symposium Chair for IEEE ICCC 2019.

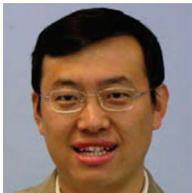

**Wei Zhang** received the Ph.D. degree in electronic engineering from The Chinese University of Hong Kong in 2005. He is currently a Professor with the School of Electrical Engineering and Telecommunications, University of New South Wales, Sydney, NSW, Australia. He has published more than 200 articles and holds five U.S. patents. His research interests include millimetre wave communications and massive MIMO. He is the Vice Director of the IEEE ComSoc Asia Pacific Board. He serves as an Area Editor for the IEEE TRANSACTIONS ON WIRELESS COMMUNICATIONS and the Editor-in-Chief for Journal of Communications and Information Networks.